\documentclass[10pt,twocolumn,floatfix,reprint]{revtex4}

\usepackage{graphicx,amssymb,amsmath}
\usepackage[utf8]{inputenc}
\usepackage{hyperref}
\usepackage{array}

\newcolumntype{L}[1]{>{\raggedright\let\newline\\\arraybackslash\hspace{0pt}}m{#1}}
\newcolumntype{C}[1]{>{\centering\let\newline\\\arraybackslash\hspace{0pt}}m{#1}}
\newcolumntype{R}[1]{>{\raggedleft\let\newline\\\arraybackslash\hspace{0pt}}m{#1}}

\begin{document}

\title{Multilevel comparison of large urban systems}

\author{Denise Pumain\footnote{Correspondence and requests for
    materials should be addressed to DP (Email:
    pumain@parisgeo.cnrs.fr).}, Elfie Swerts, Clémentine Cottineau,
  Céline Vacchiani-Marcuzzo, Antonio Ignazzi, Anne Bretagnolle,
  François Delisle, Robin Cura, Liliane Lizzi and Solène Baffi}

~

\affiliation{G\'eographie-Cit\'es, CNRS-Paris 1-Paris 7 (UMR 8504), 13 rue du
  four, FR-75006 Paris, France}

\begin{abstract}

  For the first time the systems of cities in seven countries or
  regions among the largest in the world (China, India, Brazil,
  Europe, the Former Soviet Union (FSU), the United States and South
  Africa) are made comparable through the building of spatio-temporal
  standardised statistical databases. We first explain the concept of
  a generic evolutionary urban unit (“city”) and its necessary
  adaptations to the information provided by each national statistical
  system. Second, the hierarchical structure and the urban growth
  process are compared at macro-scale for the seven countries with
  reference to Zipf’s and Gibrat’s model: in agreement with an
  evolutionary theory of urban systems, large similarities shape the
  hierarchical structure and growth processes in BRICS countries as
  well as in Europe and United States, despite their positions at
  different stages in the urban transition that explain some
  structural peculiarities. Third, the individual trajectories of some
  10,000 cities are mapped at micro-scale following a cluster analysis
  of their evolution over the last fifty years. A few common
  principles extracted from the evolutionary theory of urban systems
  can explain the diversity of these trajectories, including a
  specific pattern in their geographical repartition in the Chinese
  case. We conclude that the observations at macro-level when
  summarized as stylised facts can help in designing simulation models
  of urban systems whereas the urban trajectories identified at
  micro-level are consistent enough for constituting the basis of
  plausible future population projections.

\end{abstract}

\maketitle

\section*{}

Urban theories and models are too rarely tested on sets of data that
are properly defined and standardized. Many contradictory results and
some controversial papers in urban studies can be explained by a lack
of attention paid to the quality and quantity of empirical data. We
think of crucial importance to establish solid and replicable results
from sound data that are made comparable by using a common theoretical
background for defining and delineating cities, whatever the
heterogeneity of the published statistical information. We take here
the opportunity of several coordinated PhD works\footnote{Elfie
  Swerts, 2013, Clémentine Cottineau, 2014, Antonio Cosmo Ignazzi,
  2015, Solène Baffi, 2015} for comparing urban systems in seven among
the largest countries in the world: China, India, Brazil, Europe, the
Former Soviet Union (FSU), the United States and South Africa. This
sample includes all the so-called BRICS countries that were for a
while the most rapidly growing countries and illustrate urban systems
in almost all continents at different stages of the urbanization
process.

\section*{Standardized databases for comparing urban systems according to an evolutionary concept of cities}

Harmonised databases derived from an evolving concept of the city are
a prerequisite for comparative urban studies. In our perspective of
spatio-temporal comparisons, we define a city as a place in which the
daily activities of most residents are concentrated. Its delineation
constitutes a spatial "envelope" that evolves through time, generally
in expansion. For each time period, we select an urban delineation
that is suited to the local regime of socio-spatial interaction: thus,
before the 19th century, administrative units (communes, municipios,
or places) are sufficient to define "cities without suburbs" (see
\cite{Bairoch:1985}, p.291) that represent a dense body of population
having requested and/or having been granted legal recognition by the
political power. The form taken on by this recognition and the
conditions depend on the political and institutional setting (for
instance, request to be incorporated into the United States, or
demographic threshold in Europe, or political decision as observed
more recently in the USSR or China). With the industrial revolution,
the city became a dense human group extending outside administrative
boundaries, and it then requires a morphological definition
(i.e. urban agglomeration) so as to reflect this continuity of
built-up area. More recently, interactions linked to the use of car
transport and long-distance commuting by people living in peri-urban
settings but still working in the city centres led on to a functional
description of the city (i.e. functional urban area)
\cite{Bretagnolle:2011}.

\medskip

\begin{figure*}[!ht]
  \begin{center}
      \includegraphics[width=\linewidth]{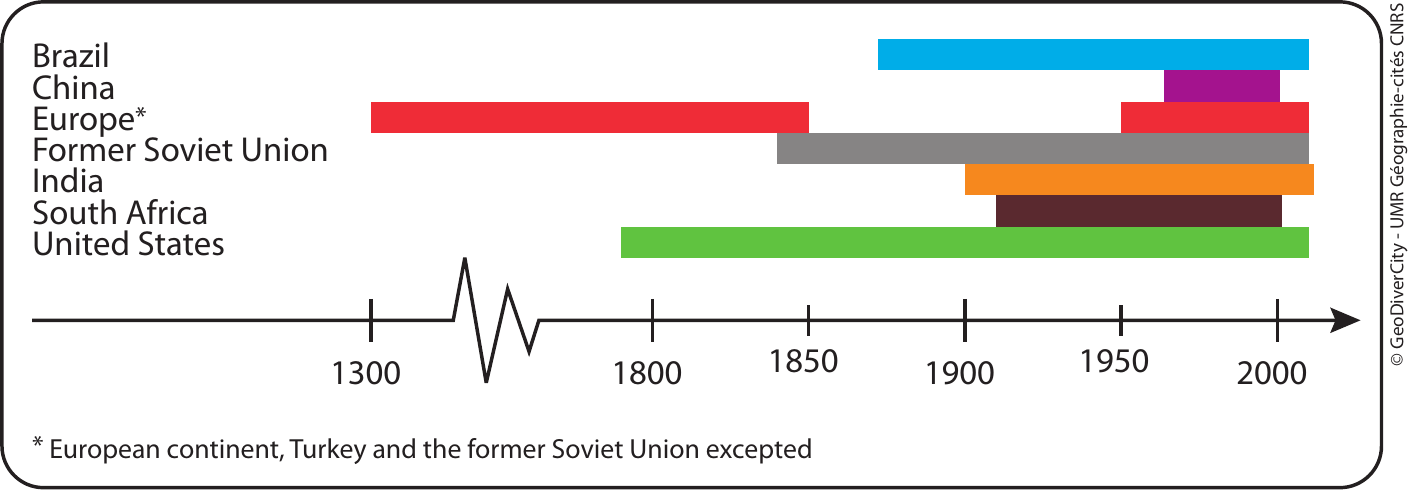}
  \end{center}
  \caption{{\bf Historical range of the urban data bases.}}
  \label{fig:historical-range}
\end{figure*}

The present development of urban databases for the seven study zones
is based on common principles, while at the same time allowing for
adaptation to local constraints and political and administrative
contexts\footnote{We would like to acknowledge the assistance of
  Hélène Mathian in the conception of the data models.}. Wherever
possible, we have used population numbers derived from the finest
possible administrative level (the basic units in the databases) and
we performed aggregations of these elementary entities in order to
follow the evolution of the urban object in the seven selected
countries over long time spans (see Figure
\ref{fig:historical-range}).

Table \ref{table:databases} sums up the final content of these
databases and the method used to adapt our generic definition to the
information available in each country. Few of these countries provide
data based on a functional definition of the urban object, and the
information is available only for certain dates (USA, Brazil). The
database for Europe, which covers a longer historical time span,
groups entities according to a morphological definition. Certain
countries only give information at administrative level, fortunately
in fairly wide units that are able to evolve (Russia:
\cite{Cottineau:2013}). Different aggregation methods have been
implemented to harmonise these databases: morphological aggregation on
the basis of aerial photographs or satellite images (Europe, India,
FSU); for South Africa, since the morphological definition lacks
relevance in the case of the Apartheid city, entities that were
functionally linked to the city, such as the black townships, were
integrated \cite{Vacchiani-Marcuzzo:2005} ; the database for the USA
uses an evolving definition of the city, considering legal entities
alone until the change to urban agglomerations in 1870 and finally
metropolitan areas (which are functional areas) from 1940
\cite{Bretagnolle:2008}; For China, the principle of urban
agglomeration was applied, adjusting delineations of built-up areas
collected on satellite images on the administrative grid of the qu,
shi and xian \cite{Swerts:2013}.

\medskip

\begin{figure}[!ht]
  \begin{center}
    \includegraphics[width=\linewidth]{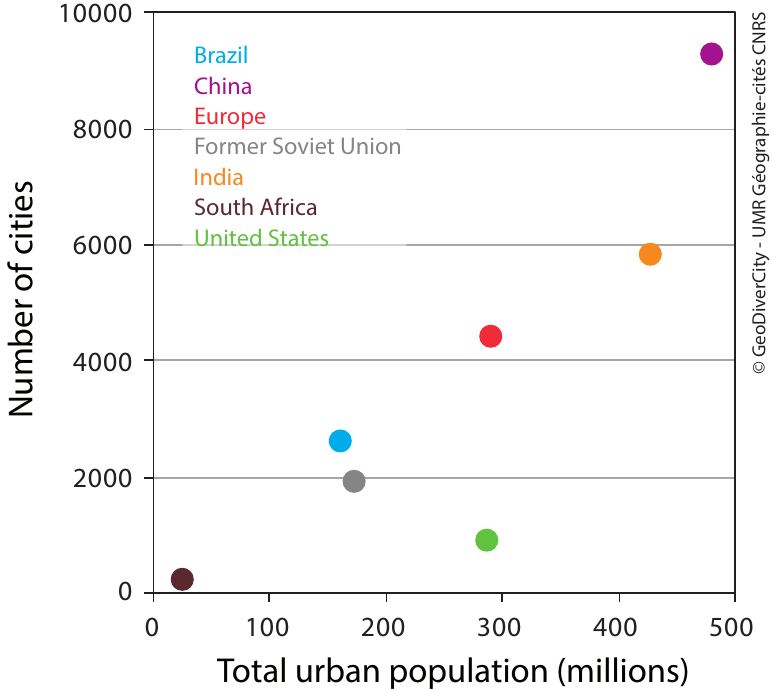}
  \end{center}
  \caption{{\bf Total urban population and number of cities per
      country.}}
  \label{fig:total-urban-pop}
\end{figure}

The urban entities included in table 1 are those with more than 10,000
inhabitants at the date indicated. At the start of the 21st century,
the numbers can vary from a few hundred to several thousands according
to the country, but they remain fairly closely linked to the total
urban population of the country, as shown in Figure
\ref{fig:total-urban-pop}, where a simple linear regression adequately
fits three quarters of the variations of these two values. Each figure
is foreseeable if the other is known, which means that the degree of
concentration of the urban population varies little in the present-day
world (a very approximate measure of the present degree of
concentration in these large countries can be obtained by calculating
an average size per city, about 100,000 inhabitants, which has of
course no real geographical meaning) and this reflects a certain
coherence in urbanisation processes overall, whatever the country
considered. The United States appear as an exception, with far less
urban units than expected from their total urban population. That huge
concentration of the urban system can be explained by the historical
process of settlement in the “New world” as well as by the very large
size of the elementary spatial units aggregated in SMA’s (the
counties). Even if we had included in the database the urban
agglomerations instead of the SMA’s, the exception would remain, since
there are 1380 “urbanized areas and urban clusters” over 10 000
inhabitants for a total of 237 million urban population (US Census
Bureau, 2014)\footnote{Island Areas, Alaska and Porto Rico excluded}.

\begin{table*}
  \centering
  \begin{tabular}{C{3cm}|c|c|c|C{8cm}}
    Database & $N_0$ (initial date) & $N_F$ (final date) & $D$ & Method \\
    \hline
    Brazil (a) & 531 (1872) & 2615 (2010) & 11 & Administrative
    (municipios + metropolitan areas) \\
    \hline
    Former Soviet Union (DARIUS) (b) & 91 (1840) & 1929 (2010) & 11 &
    Morphological (municipalities at three administrative levels) \\
    \hline
    India (IndiaCities) (c) & 503 (1901) & 5841 (2011) & 12 & Combined
    morphological, and functional \\
    \hline
    China (ChinaCities) (d) & 605 (1964) & 9294 (2000) &  4 & Combined
    administrative (xian, qu, xianjishi (district level) and zhen,
    xiang (subdistrict level)) morphological and functional \\
    \hline
    South Africa (DYSTURB) (e) & 14 (1911) & 220 (2001) & 10 & Combined
    morphological and functional (white cities + black townships) \\
    \hline
    Europe (PARIS-Bairoch-Geopolis) (f) & 3619 (1950) & 4413 (2010) & 14 &
    Morphological (municipalities then agglomerations) \\
    \hline
    United States (Harmonie-cités) (g) & 5 (1790) & 909 (2010) & 23 & Combined
    administrative then morphological and functional (places, cities,
    retropolated micropolitan and metropolitan areas) \\
    \hline
  \end{tabular}
  \caption{
  	\textbf{Harmonised urban databases for international comparisons (urban units $>$ 10,000 inhab.).}
    $N_0$ : number of cities at (initial date) ; 
    $N_F$ :  Number of cities at (final date) ; 
    $D$ : number of dates considered in the database.\\
     Sources : (a) IBGE, Instituto Brasileiro de Geografia e Estatistica. \href{http://www.ibge.gov.br/home/mapa\_site/mapa\_site.php\#populacao}{Statistical tables for the Brazilian Empire and the Federal Republic of Brazil} ; 
     (b) \href{www.demoscope.ru}{Statistical tables} for the Russian Empire, Russian Empire censuses, USSR censuses (archives), National censuses ;
     (c) Census of India ;
     (d) China Data Center and Chinese National Census Bureau ;
     (e) Statistics South Africa,  South African institute of Race relations, Urban foundation, Census Statistics SA, \href{http://www.cartographie.ird.fr/dysturb.html}{DYSTURB} (see \cite{Giraut:2009}) ;
	 (f)  \cite{Bairoch:1985, Moriconi:1994} ;
	 (g) \href{http://www.census.gov/prod/www/abs/decennial/}{USA census data: Census of Population and Housing}.
}
  \label{table:databases}
\end{table*}

\section*{Forms of urban hierarchy and growth processes}

There is a vast corpus of literature about the forms taken on by urban
hierarchies, and in particular the descriptions that refer to Zipf’s
rank-size rule. Recent syntheses \cite{Nitsch:2005, Schaffar:2012}
have not reached a consensus on the universal nature of this rule, nor
on the factors that might explain its wide applicability, or the main
reasons for the variations that are observed.

A first obstacle to the construction of scientific knowledge on this
issue is the extreme heterogeneity of the samples of cities that have
been used to perform the tests, and sometimes the doubtful quality of
the definitions and delineations used to measure city size. We do not
claim here to provide a final solution to this problem, but we do
bring more credible results by using databases that are as comparable
as possible, applying this to a large number of cities over fairly
long time spans and for a variety of large countries of the world.

Another question is that of the definition of the city systems
observed. Most often, Zipf’s law is tested on a set of cities in a
single country. We are aware that this use of the nation-state
framework (or a quasi continental framework in the case of Europe, US,
China, India and the former Soviet Union) is probably no longer
completely suited to the delineation of city systems, since the
globalisation of exchanges brings cities to new interactions, the
intensity and range of which vary according to city
size. Nevertheless, urban hierarchical patterns, precisely because of
the growth processes resulting from their interaction patterns, tend
to be sustained over periods of several decades or even several
centuries, and we think that the hierarchies observed here have been
engaged in strong interactions for sufficiently long time for the
patterns to continue to show up, even if the state borders enclosing
them are no longer as impermeable as they were earlier.

\subsection*{Macro-level analysis of urban hierarchies explained by territorial history}

Our main results widely confirm the results reported by
Moriconi-Ebrard \cite{Moriconi:1993} who used the Geopolis database to
compare states across the world. They also go against the idea of a
historical convergence towards a regular Zipf model with a slope of -1
as suggested in a recent article by Berry and Okulicz-Kozaryn
\cite{Berry:2012}.

\medskip

Indeed, slope values for rank-size adjusted on these distributions
vary, clearly differentiating countries and continents according to
how long-standing their settlement is. The value of the rank-size
slope is an index of size inequalities of cities. The variations of
this index of urban size amplitude in a given territory can be fairly
well explained by differences in the speed of transport systems
enabling exchanges among cities, at the time when the urban networks
become established: the lowest values (absolute value, i.e. without
sign) are observed in countries that have been populated for a long
time (India, China, Europe) and the highest values for countries that
were settled more recently, higher transportation speed enabling a
wider spacing between settlements as well as larger urban
concentrations emerging on sparser spatial distributions of rural
population (South Africa, the United States). The fairly high values
for the Soviet Union could be explained by its relatively late
industrialisation compared to Europe, and a more recent urban
development of its Eastern areas in the Asiatic part, while Brazil
escapes the general pattern with a moderate degree of urban
concentration (see Table \ref{table:urban-hierarchies}).

\begin{table*}
  \centering
  \begin{tabular}{cccccc}
    Country & $N$ & $a$ & $P_1/P_2$ & $M$ & $P_{tot} (\times 10^6)$\\
    \hline
    Brazil & 2615 & 0.88 & 2 & 2 & 161 \\
    FSU & 1929 & 1.10 & 3 & 0 & 173.5 \\
    India & 5121 & 0.95 & 1.1 & 3 & 427 \\
    China & 9294 & 0.80 & 1.3 & 0 & 481 \\
    South Africa & 220 & 1.15 & 2 & 4 & 25 \\
    Europe & 4413 & 0.96 & 1.2 & 2 & 291 \\
    United States & 909 & 1.23 & 1.5 & 0 & 287\\
    \hline
  \end{tabular}
  \caption{{\bf Comparing urban hierarchies: city size distributions around 2010.} 
    We consider the urban agglomerations larger than $10000$ inhabitants. 
    Rank size slope $a$ is estimated from equation $log(P) = K - a
    log(R)$ with $P$ the population of the city and $R$ its rank,
    using OLS method. $P_1/P_2$: primacy index; $M$: number of
    macrocephalic cities; $P_{tot}$: total urban population.
  }
  \label{table:urban-hierarchies}
\end{table*}

\begin{figure}[!ht]
  \begin{center}
    \includegraphics[width=\linewidth]{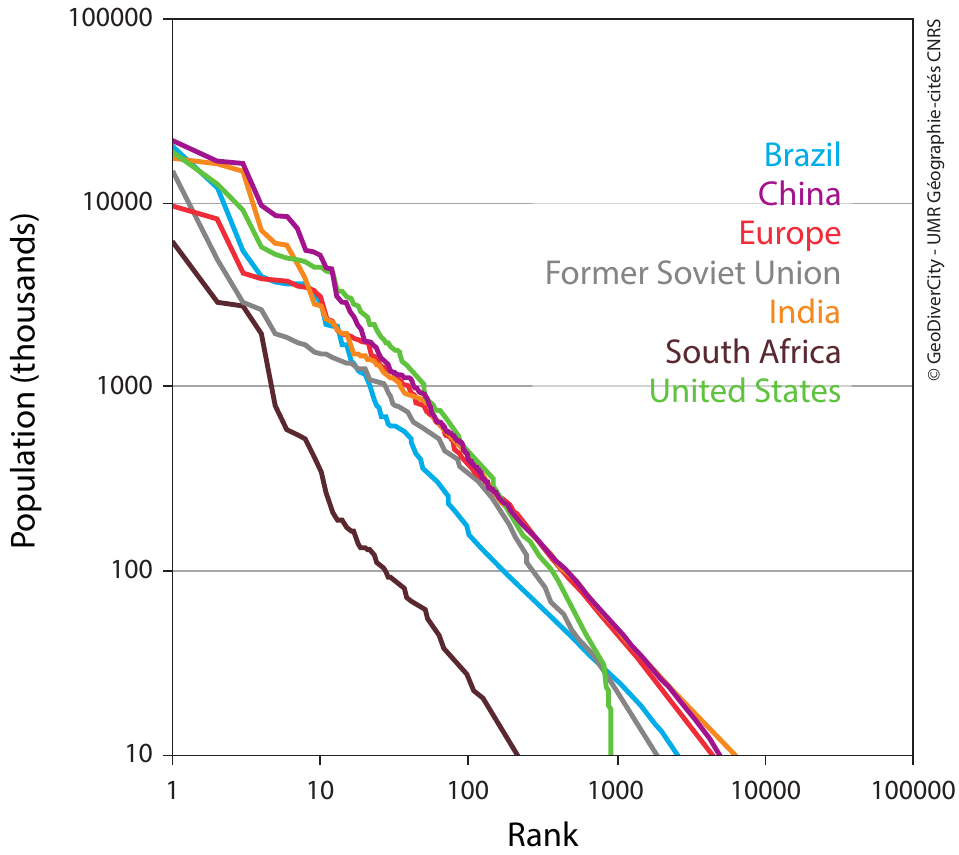}
  \end{center}
  \caption{{\bf City size distributions in seven countries around
      2010.}}
  \label{fig:city-size-distributions}
\end{figure}

The qualitative variations in shape of the size distribution are
explained above all by the diversity of the politico-administrative
organisation of the territories concerned (see Figure
\ref{fig:city-size-distributions}). In countries that have been run under
socialist regimes aiming at restraining urban growth, there is a
levelling-off of the curves (in the FSU around one million
inhabitants, in China around 100 000 inhabitants, corresponding to
city sizes for which targeted investments have been made
\cite{Clayton:1989, Moriconi:1993}. Conversely the countries with the
most marked macrocephaly (South Africa, India, Brazil) are those that
have allowed their metropolises the greatest latitude.

Compared to the United States and Europe, the BRICS countries stand
out for the particular shape of the upper part of their urban
hierarchies: Russia is the only BRICS country to present a case of
urban primacy, where Moscow is three times the size of St Petersburg
(see Table \ref{table:urban-hierarchies}) while all the others except
China are characterised by macrocephaly comprising two to four cities
that are clearly discontinuous with the rest of the distribution. Thus
in India, the three cities of Delhi, Mumbai and Kolkata, with around
16 million inhabitants each, stand out from the other cities in India,
as do Durban, Cape-town and Johannesburg agglomeration in South
Africa, while in Brazil the discontinuity is less marked, with Sao
Paulo (20 million inhabitants) and Rio de Janeiro (12 million) some
way ahead of Belo Horizonte with just over 5 million.

The largest cities are remarkable by their size, which obviously
depends on the mass of the urban population in the country (see Figure
\ref{fig:total-urban-pop}), but this top part of the urban hierarchy
is not sufficient to characterise the degree of concentration of an
urban population, and it is relevant to look at the other parts of the
distribution. Thus China, which ranks first for its mass, with a total
urban population of around 500 million inhabitants, also ranks first
for the number of cities of over a million inhabitants, 66 in all. In
contrast, India, ranking second for the weight of its total urban
population (around 400 million) ranks only 4th for the number of
cities with more than a million inhabitants – with 44 of these, it is
outranked by the USA which has 51 for an urban population of only 287
million, which is therefore much more concentrated spatially. In
contrast again, Europe where the urban population is more or less the
same than in the USA (291 million) has only 39 cities of over a
million inhabitants, and the degree of concentration is fairly similar
to that of the former Soviet Union (173.5 million city-dwellers and 28
cities of over a million), or Brazil (161 and 23 respectively). While
Europe remains the continent of small to medium cities and towns, and
while India still has many small cities, the urban processes underway
in India suggest that many of these cities are set to expand in the
coming decades \cite{Swerts:2013b}, and in China, although population
concentration is still moderate, the massive size of the country and
the closeness of certain cities one to the other suggests that several
large conurbations or megalopolises are likely to develop with 30 to
40 million inhabitants each, around the Pearl River Delta, in the
regions round Shanghai, or between Beijing and Tianjin.

\subsection*{Urban hierarchies and urban growth at micro-level: testing Gibrat’s model}

The universal shape of urban hierarchies that can be summarized by
Zipf’s rank-size rule as above or by a lognormal statistical
distribution is explained as a first approximation by a stochastic
repartition of the growth rates of individual cities in an urban
system \cite{Robson:1973, Pumain:1982}. Gibrat’s law
\cite{Gibrat:1931} predicts the statistical form of urban hierarchies
and their persistence over time. This model assumes that cities grow
in a manner that is proportional to their size. It is a growth model
of the exponential type, but with growth rates (or relative population
variations) that vary in the course of time. According to this model,
despite considerable fluctuations in growth rates from one city to
another over short time spans, long-term growth averages out at the
same level for all cities in a given system.

We tested the hypotheses of Gibrat's law for the intervals between
these two dates for which city population data is known in our
bases. The hypothesis for generating a lognormal distribution
stipulates that the variations in growth rate at each time interval do
not depend on city size, and are distributed randomly from one period
to another. Overall, the process observed in the BRICS complies with
the model, which thus, at least in first approximation, remains a
relevant reference for analysing the process of growth distribution in
the city systems.

Everywhere, the correlation between city size and growth is low or
absent. In South Africa, in India and China, whether at national level
or for the main regions, the hypotheses of the Gibrat model are
verified, and from the start of the 20th century
\cite{Vacchiani-Marcuzzo:2005, Swerts:2013b}. There is a discrepancy
with the model, particularly in periods of vigorous growth, for the
USA throughout the 19th century and in Europe after 1950, where there
is a positive temporal autocorrelation of the growth rates. In Russia
the process also appears in the course of the second phase of
industrialisation in the 1930s and in the two decades during which
there was a trend towards metropolisation of the largest cities and a
cumulative decline of certain specialised cities
\cite{Cottineau:2013}. We have demonstrated in a previous paper
\cite{Favaro:2011} how such deviations from a purely stochastic growth
model could be explained by integrating in an urban growth model the
interaction processes that convey innovation waves in the urban
hierarchies. A consequence is that inequalities in city sizes are
growing faster in real urban systems than according to a pure Gibrat’s
rule.

To conclude, the urban hierarchies and growth processes in a variety
of large urban systems all over the world, including the BRICS
countries, share generic common features despite major differences in
their history as well as territorial and political organization. That
is why Zipf’s law and Gibrat’s model albeit purely statistical models
remain rather good standard references enabling international
comparisons for a synthetic description of empirical urban hierarchies
and urban growth processes, even though they do not directly provide
an explanation for the underlying generative geographical processes
(we suggest further theoretical complementary investigations in this
direction, for instance by developing the family of Simpop models,
\cite{Pumain:2012, Cottineau:2015}. This conclusion is consolidated by
the consistency of deviations from the models that can be rather
easily related to different families of the historical development of
urban settlements and politico-administrative organisation of the
territories they belong to.

\section*{Macro and micro-dynamics: urban transitions and cities trajectories}

As we have seen before, understanding city size distribution and
moreover urban growth processes cannot be done without referring to
the history of each urban system and at least, even in a parsimonious
abstract approach, to its stage in the seemingly universal urban
transition. This process named by W. Zelinsky \cite{Zelinsky:1971}
using an analogy with the demographic transition describes the
universal change from a dispersed and homogeneous spatial repartition
of population in rural habitat towards much more concentrated
diversified and hierarchized forms in urban settlements. The process
which generally accompanies economic development started roughly at
the beginning of 19th century in first industrialized countries and
around 1950 in the less developed ones.

\subsection*{Macro-level trajectories of urban systems}

Different stages in urban transition clearly appear already from the
condensed information in Table \ref{table:avg-annual-growth-rates}
where countries that experimented earlier urban transitions (Europe,
Former Soviet Union, USA, Brazil) have much lower average urban growth
rates during the last forty years than countries that are still in the
exponential stage of the growing curve of their urbanization rate.

\begin{table*}
  \centering
  \begin{tabular}{ccc}
    Country & Average growth rate ($\%/year$) & Period \\
\hline
Brazil & 1.11 & 1960-2010\\
China & 5.20 & 1964-2000\\
India & 2.10 & 1961-2001\\
Former Soviet Union & 0.78 & 1959-2010\\
South Africa & 3.15 & 1960-2001\\
Europe & 1.01 & 1960-2010\\
USA & 1.52 & 1960-2010\\
\hline
\end{tabular}
\caption{{\bf Average annual growth rates of population for cities of over $10.000$ inhabitants during second half of the 20th century.}}
  \label{table:avg-annual-growth-rates}
\end{table*}

China appears clearly as the country in which recent urban development
is the most rapid, with an average urban population growth rate of
over 5\% per year over 40 years. The Indian urban growth rate is half
of this, but nevertheless vigorous during the period, with an average
rate of 2\% per year, while South Africa, with 3.2\%, still reflects
the fast urban growth rates across the African continent. In the USA
and in Brazil growth is still over 1\%, while in Europe and Russia it
is just 1\% or much lower.

\medskip

These trends are partly determined by the stage reached by the
different countries in the urban transition, but they are also
influenced by their particular history of urbanisation: in China
political action, in particular that affecting migrations, for a long
time put brakes on the explosion of urbanisation \cite{Chan:1985,
  Lin:2002}, far more markedly than in India where it was rather
social and family ties that slowed migration from rural areas
\cite{Banerjee:1981, Ramachandran:2011}. South Africa has remained at
an average level among African countries since the end of Apartheid,
on the one hand because of internal migrations from the former
Bantustans, and on the other because of its attractiveness towards
foreign migrants \cite{Davies:1986, Vacchiani-Marcuzzo:2005}. Brazil
and the USA being among the "new" countries in terms of waves of
settlement had a former urbanisation rate systematically higher than
in "old world" countries in Europe and Asia \cite{Santos:2008}. The
slowing in urban growth rates has been more marked in Russia since the
1990s, as a result of decreases in the total population, especially in
the Northern parts of the territory resulting from the dismantling of
the Soviet Union \cite{Lappo:1999, Eckert:2003}. It has even been
negative on average over the last two censuses (since 1989).

\medskip

\begin{figure}[!ht]
  \begin{center}
    \includegraphics[width=\linewidth]{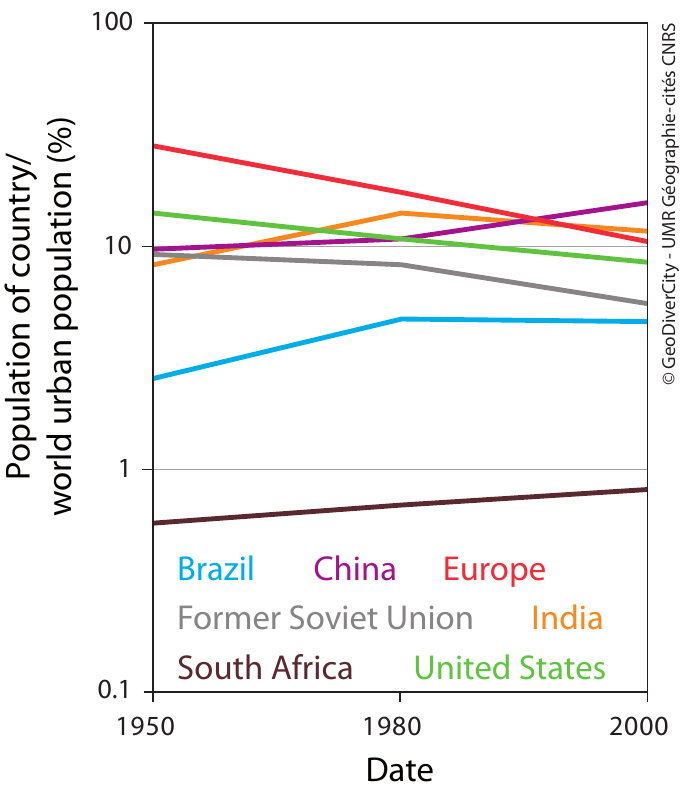}
  \end{center}
  \caption{{\bf Comparing the evolution of countries share of world
      urban population (\%).}}
  \label{fig:coutries-share}
\end{figure}

As a result, the weights of urban population of these countries in the
total urban population of the world have been relatively increasing as
shown in Figure \ref{fig:coutries-share} that compares their evolution
over almost half a century.

These contrasted evolutions explain the very rapid turnover in
rankings among the mega-cities (above 10 millions inhabitants), which
are always the subject of controversy because the delineations chosen
to measure these large urban areas have a considerable impact on their
rank\footnote{See for instance the classifications given on websites
  http://citypopulation.de, http://population.data.net, or those
  provided by the United Nations (http://esa.un.org/unup).}. We can
consider here the subset of these seven large countries included in
the harmonised database to demonstrate the upheavals caused by unequal
demographic growth over the last four decades. In the 1960s, the three
largest cities by size were located outside the BRICS, with
populations of 10.6 million for New York, 8.9 for Greater London, and
7.2 for Paris. These were followed on by Moscow (7.2), Shanghai (6.4),
Delhi (5.9), Los Angeles (6), Kolkata (5.3), Mumbai (4.9), Beijing
(4), Sao Paulo (with 3.8 having just overtaken Rio at 3.3) and
Guanzhou (2). By the end of 20th century, this ranking had been
completely overturned, with Sao Paulo joining or overtaking New York
with some 20 million inhabitants, and 4 cities in the BRICS taking top
places among the others: Delhi (17 million), Shanghai (16), Kolkata
(16), Mumbai (15), Moscow (14), Beijing (14), and Rio (12). The
European cities do not exceed 10 million, and they are joined in this
megapolis group by Guanzhou, Shenzen, and probably a few other Chinese
cities if the migrant populations with more or less illegal status are
taken into account.

\subsection*{Differentiating urban trajectories at micro-level}

Too often however in urban studies the focus is only directed on these
few “global cities” or “world cities” whose size and expansion is
linked with the development of long distance exchange networks but not
necessarily reflect other possible types of urban dynamics. We think
necessary to develop knowledge about perhaps less prestigious names in
urban hierarchies that nevertheless provide ways of living for a
majority of populations and also participate in a decisive manner to
the maintenance and renewal of urban systems.

\begin{figure*}
  \begin{center}
    \includegraphics[height=\textheight]{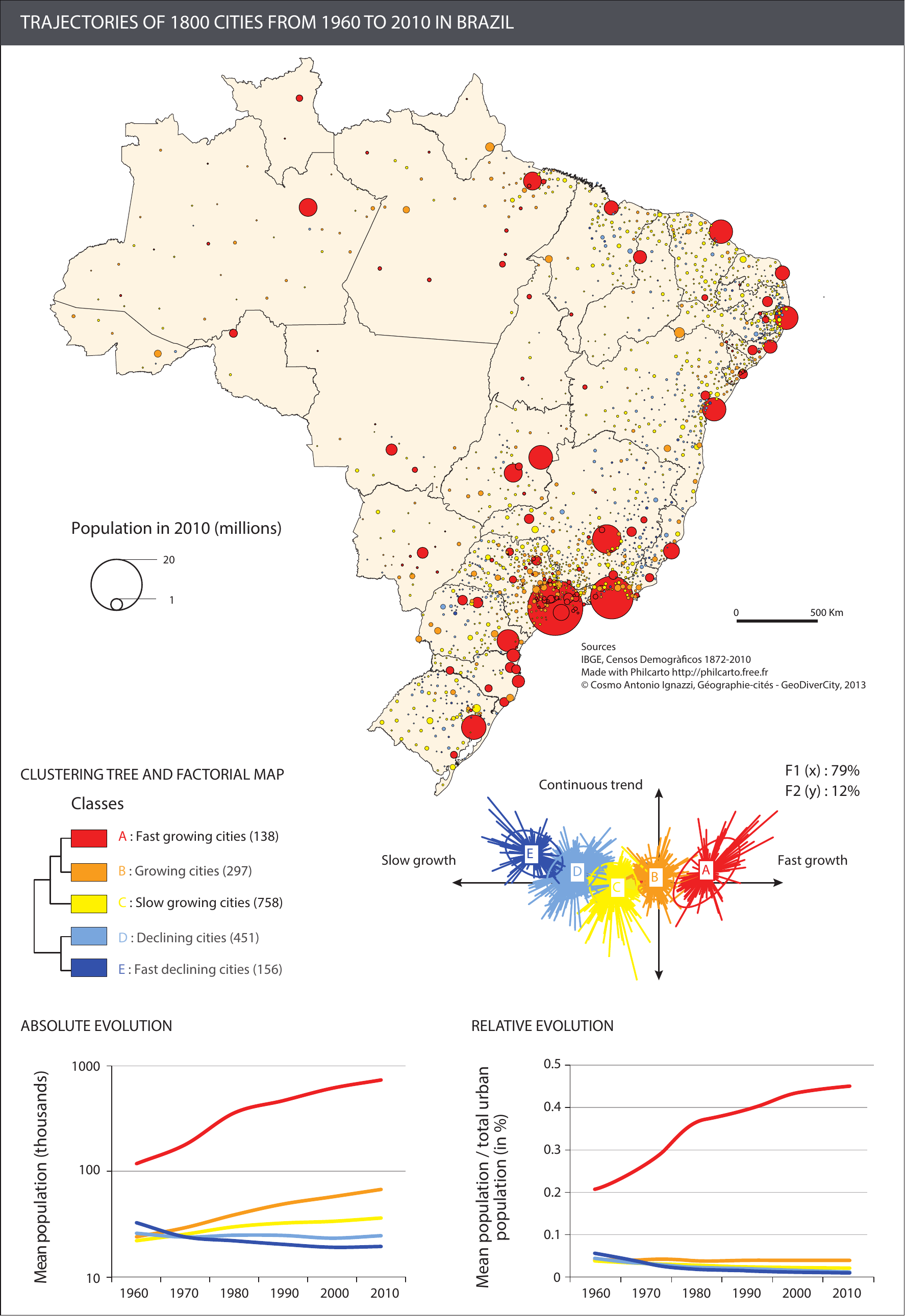}
  \end{center}
  \caption{{\bf Urban trajectories in Brazil.}}
  \label{fig:brazil}
\end{figure*}

\begin{figure*}
  \begin{center}
    \includegraphics[height=\textheight]{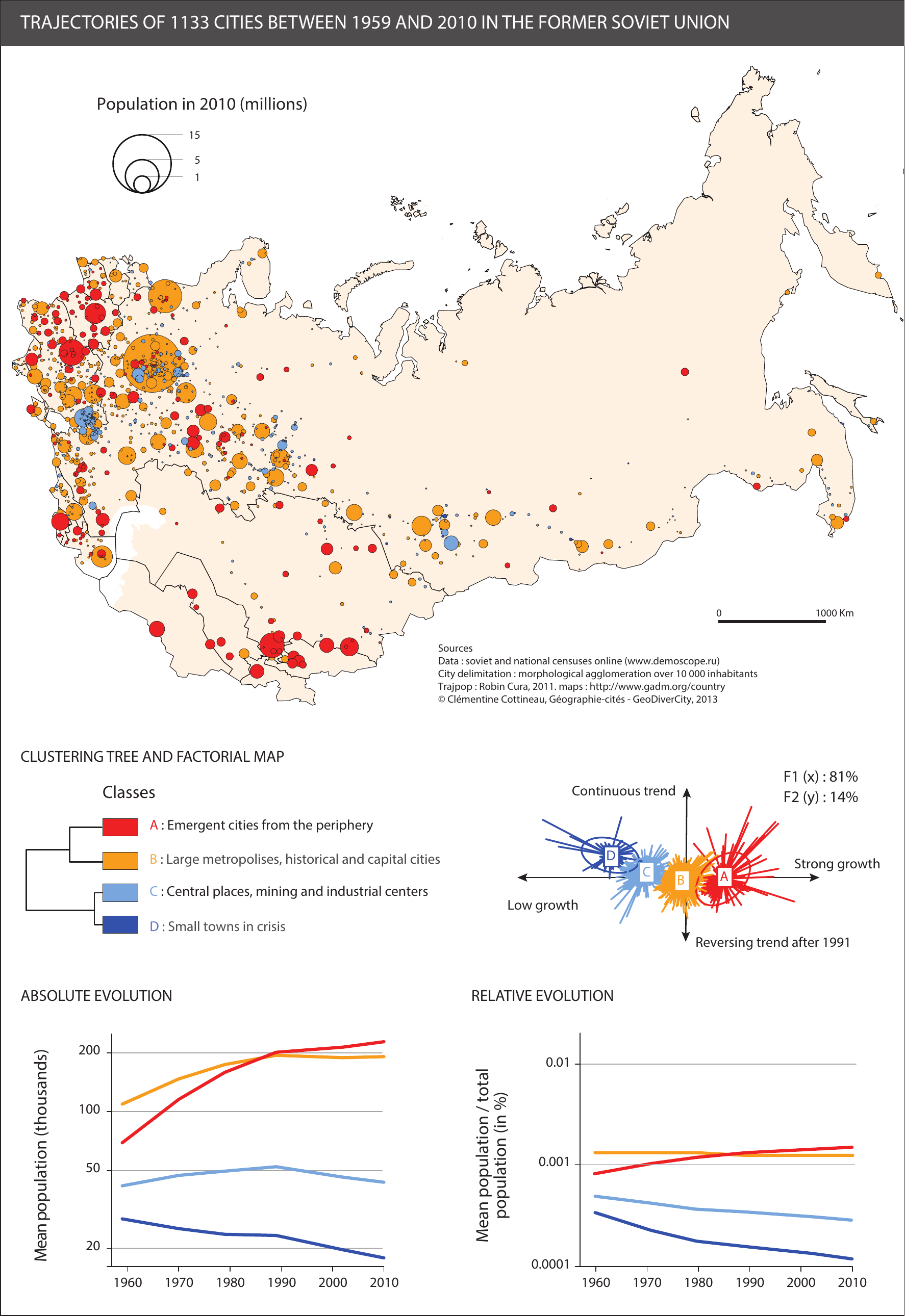}
  \end{center}
  \caption{{\bf Urban trajectories in former Soviet Union.}}
  \label{fig:soviet-union}
\end{figure*}

\begin{figure*}
  \begin{center}
    \includegraphics[height=\textheight]{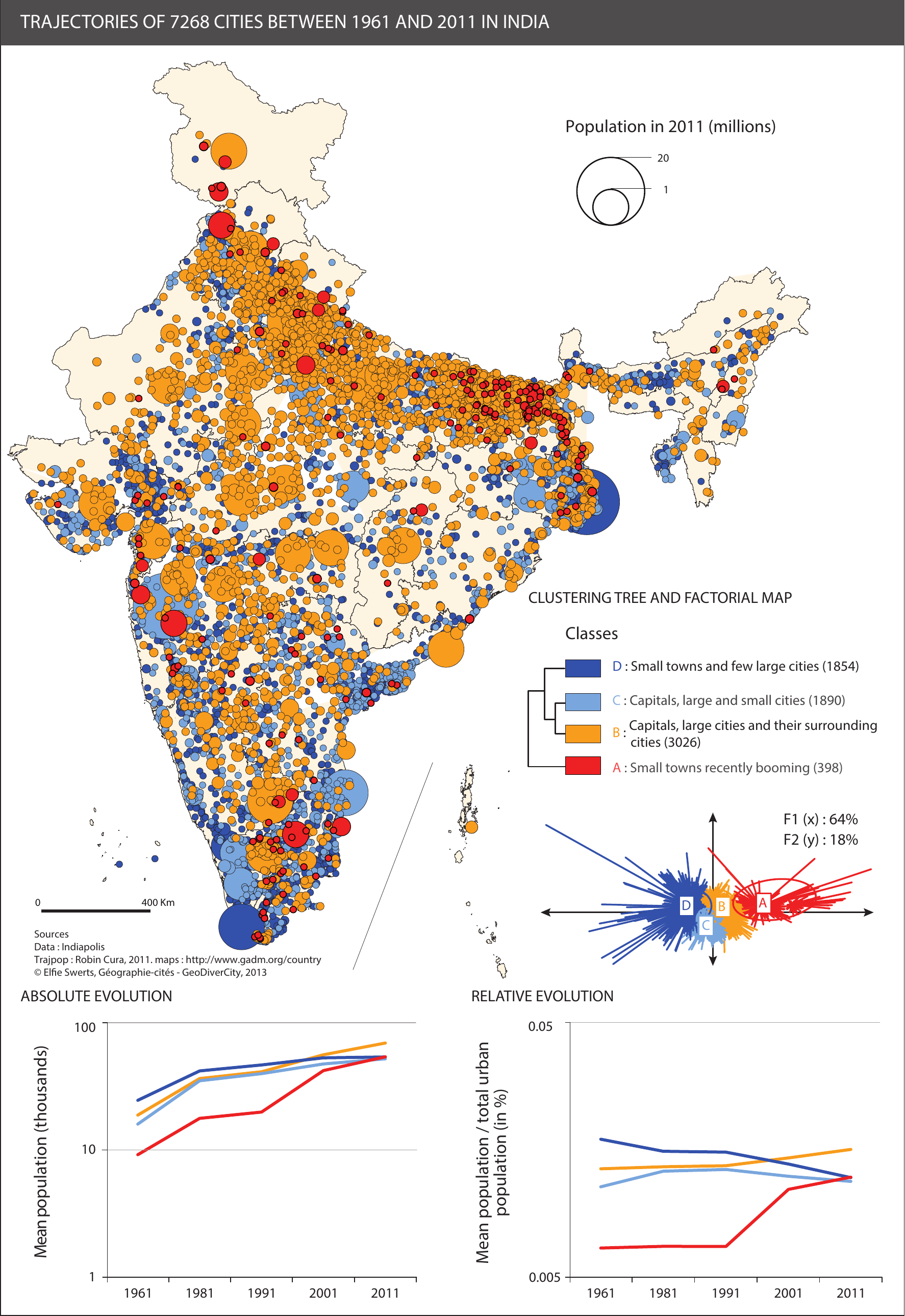}
  \end{center}
  \caption{{\bf Urban trajectories in India.}}
  \label{fig:india}
\end{figure*}

\begin{figure*}
  \begin{center}
    \includegraphics[height=\textheight]{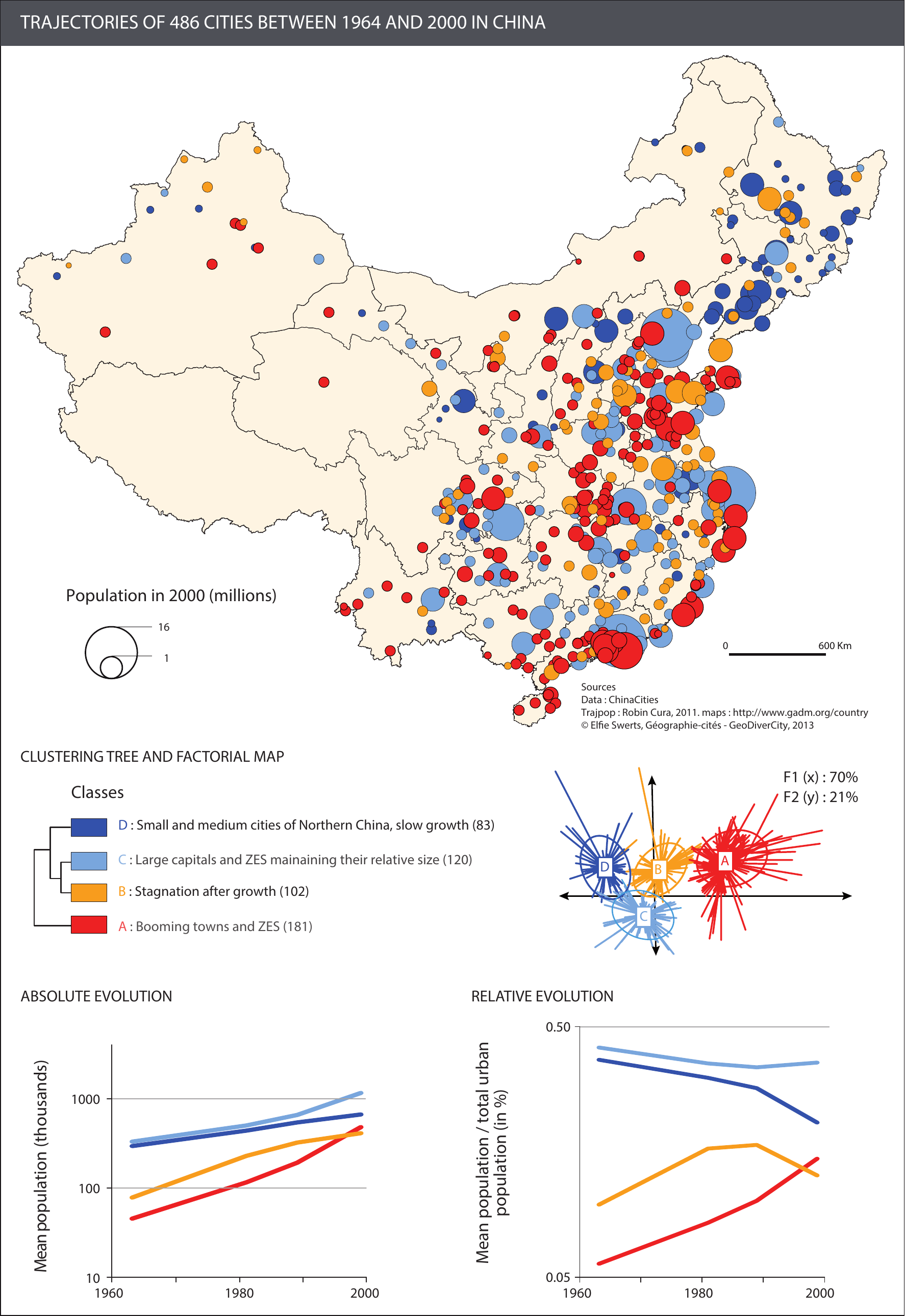}
  \end{center}
  \caption{{\bf Urban trajectories in China.}}
  \label{fig:china}
\end{figure*}

\begin{figure*}
  \begin{center}
    \includegraphics[height=\textheight]{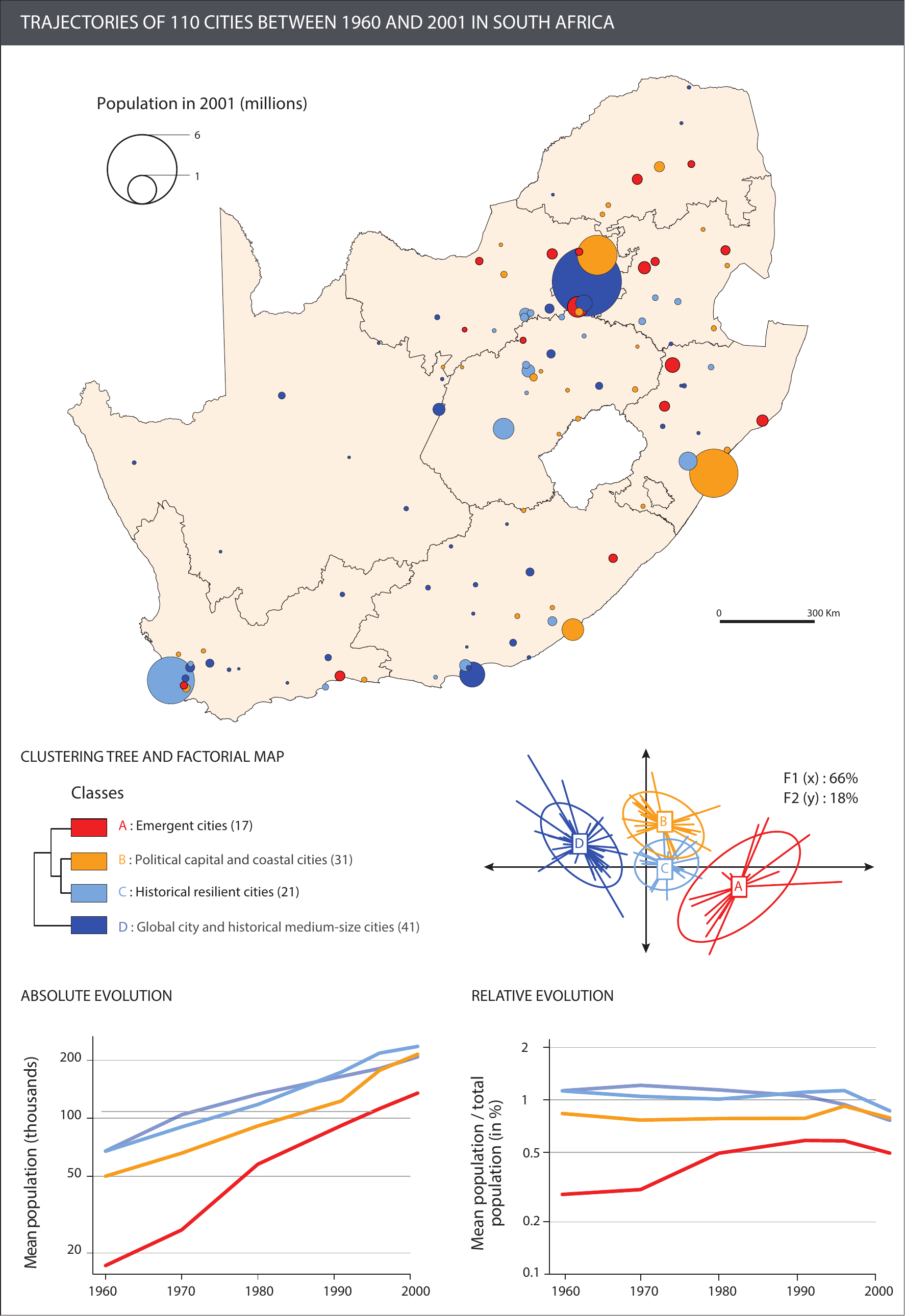}
  \end{center}
  \caption{{\bf Urban trajectories in South Africa.}}
  \label{fig:south-africa}
\end{figure*}

To differentiate city trajectories in each country we developed a
method that compares population evolution profiles by way of a
correspondence factor analysis and a hierarchical ascending
classification using the $\chi^2$ distance (see \hyperref[sec:trajpop]{Appendice 1}). The
implicit reference model is therefore that of proportional distributed
growth proposed by Gibrat \cite{Gibrat:1931}, and the trajectories
show the most systematic discrepancies in relation to mean growth
trend. The types of profile derived from the classification are shown
on the left-hand graph (see Figures \ref{fig:brazil}, \ref{fig:soviet-union},
  \ref{fig:india}, \ref{fig:china}, and \ref{fig:south-africa}) by the trajectory of the
mean population for each type, and to the right for a trajectory
showing the relative evolution of the mean weight of this type of city
in the urban system under consideration (semi-logarithmic graphs
enabling on the one hand a comparison of growth intensities
represented by the slopes, and on the other immediate visualisation of
any differences in evolution (they are fairly frequent) associated
with positions in the urban hierarchy).

Naturally, the shape of city trajectories in relation to that of the
system to which they belong can vary with the period considered. We
chose the period 1960-2010 for reasons of comparability between
countries, and also to restrict the number of major historical turning
points which if too numerous would make it more difficult to
efficiently anticipate possible evolutions in the 21st century. By
construction, these classifications show classes of cities where the
"absolute" evolutions are often all increasing (rarely including an
inflexion), but grouped according to their relative growth, which may
be faster or slower than that of the country as a whole. To compare
the degree of heterogeneity of these trajectories across countries, we
measured the share of variance remaining between classes. The order of
the countries remains the same, whether the partition is into two,
three or five classes. In the case of five classes, the Former Soviet
Union shows the greatest diversity in trajectories (76\% interclass
variance) while in Brazil the trajectories appear less differentiated
(63\%), the other countries falling between these two (70-73\%).

Depending on the form of the classification tree, there are two
clearly distinct types of trajectory in China, while three families of
trajectory can be observed in the other countries. To obtain more
detail, and in relation to levels of heterogeneity, for mapping
purposes (Figures \ref{fig:brazil},\ref{fig:soviet-union}, \ref{fig:india},
  \ref{fig:china}, \ref{fig:south-africa}) we retained four classes of city for
China (Figure \ref{fig:china}), India (Figure \ref{fig:india}), South
Africa (Figure \ref{fig:south-africa}) and the former Soviet Union
(Figure \ref{fig:soviet-union}), and five for Brazil (Figure
\ref{fig:brazil}). The classes can be grouped according to the
orientation of their trajectories in relation to the city system to
which they belong, generally in two types, "winners" and "losers", but
a stable type also appears in India, China, the former Soviet Union
and South Africa. These cities that maintain their relative weight in
the system are often long-standing cities with administrative
functions – certain State capitals in India, provincial capitals in
China regional capitals in Russia, and medium-sized cities in South
Africa.

In Brazil (Figure \ref{fig:brazil}) almost all the large metropolitan
areas, which are the capitals of the federal States, have strongly
ascending trajectories. The recent dynamic thus has the effect of
accentuating the hierarchical inequalities in the country. India too
exhibits this process of reinforcement at the top of the urban
hierarchy, with three quarters of the largest cities exhibiting
ascending or stable trajectories. In contrast, in the other countries,
what can be seen is a form of "catching up" by the smaller cities and
peripheral areas. The markedly ascending trajectories tend to be
characteristic of a few smaller cities, often in the vicinity of the
large metropolises, in India (Figure \ref{fig:india}) or South Africa
(Figure \ref{fig:south-africa}). In the former Soviet Union (Figure
\ref{fig:soviet-union}), it is most of the cities located on the
peripheries of Russia to the South (Central Asia, Azerbaijan) and West
(Ukraine, Belarus), and some Russian cities near deposits of mineral
resources, that gain weigh relative to the others. In China (Figure
\ref{fig:china}), two processes are seen, where the large cities in
the East (Shenzen, Xiamen) are gaining weight in the system while
medium-sized cities in Xinjian province and Inner Mongolia are
developing fast, illustrating the catching-up by peripheral regions.

There is a long-term trend in most city systems whereby it is mostly
the small urban entities that show relative decline. Indeed, all other
things being equal, the smaller cities are more likely to be distant
from the main waves of innovation, or else to be highly specialised in
declining sectors of activity, so that they lose their influence on
local markets as a result of acceleration in the speed and capacity of
transport systems. The only exception appears in China (see Figure
\ref{fig:china}) where classes of cities with an ascending profile are
made up mainly of small cities, a third of which in Special Economic
Zones in which innovating activities have been set up and to which
populations migrate. The classic process of hierarchical diffusion of
innovation is partially disconnected here from the previous structure
of the city system. However cities with a "winning" trajectory are
mainly located in the immediate vicinity (roughly less than 200 km) of
large metropolitan areas, for instance Guangzhou or Shanghai. The
importance of policy in urban dynamics can thus be seen in the
creation of new cities, at the same time preserving a degree of
spatial and historical coherence with the earlier trends in the city
system. In a territory where urbanisation is long-standing, like
China, these new urban developments fit themselves into the previous
urban spatial pattern, while in "new" countries like Brazil, the USA
and South Africa, and also the Eastern part of the Russian Empire,
urban creations ran alongside the settlement of new territories.

\section*{Conclusion}

When dealing with complex systems, it is important to relate the
configurations of urban hierarchies observed on macro-geographical
scale of States to the trajectories of the urban entities they
comprise on micro-geographical level. This paper thus demonstrates the
usefulness of constructing a harmonised database enabling the
description of the evolution of urban entities and their spatial
extension over time.

We provide for the first time a comparable overview of the systems of
cities in the five BRICS countries. Using Zipf’s distribution of city
sizes and Gibrat’s urban growth models as benchmarks for the
comparison, we have demonstrated that the dynamic urban processes in
BRICS during the last fifty years were rather similar to those
observed for instance in Europe or the United States. There is nothing
resembling a specific urban dynamic in BRICS whether we consider the
shape of urban hierarchies, city size distribution, or distribution of
urban growth among individual cities.

Of course differences do exist, but they relate to the specific
developmental pathway of these countries, including the relative delay
in the urban transition compared to more developed countries, which
explains their very high mean urban growth rates – the case of Russia
being excepted. History matters too for differentiating the evolution
of urbanisation rates, which registered higher values earlier in
Russia and Brazil compared to South Africa, China and India.

When shifting from the macro-scale of countries to the micro-scale of
individual cities, the most striking fact is the diversity of urban
trajectories that exhibit contrasted patterns of booming growth or
relative decline everywhere. Moreover, these qualitatively divergent
local evolutions are disseminated in all parts of each territory with
few remarkable spatial concentrations.

\section*{Appendice 1. TrajPop software (author:Robin Cura)}
\label{sec:trajpop}

These analyses are performed using the TrajPop script, developed in
the ERC GeoDiverCity project. This tool, based on the free statistical
environment R, performs a Correspondence Analysis on a temporal
population table. The coordinates of cities on the orthogonal
components then make it possible, after re-entering the weights of the
cities, to generate a matrix for population discrepancies among cities
(measured using a khi2 distance); to this matrix is applied a
Hierarchical Cluster Analysis (using the Ward method, which tends to
minimise intra-class variance and to maximise inter-class
variance). From the tree generated by this clustering, the number of
clusters is chosen so that it sufficiently distinguishes the
trajectories while at the same time enables them to be mapped. It is
then possible to analyse trajectory classes using the TrajPop graphic
and numerical printouts, for instance by studying the evolution of the
relative weights of the classes in the system in the course of
time (\url{http://trajpop.parisgeo.cnrs.fr}).

\section*{Acknowledgements}

The authors acknowledge funding from the European Research Council
through project GeoDiverCity (advanced grant programme, number 269826, PI
Denise Pumain).

\end{document}